\documentclass[epj]{svjour}
\usepackage{graphics}
\def\beq{\begin{equation}}

\def\eeq{\end{equation}}

\begin{document}
\title{Gravitons and pions}
\author{John F. Donoghue 
}                     
%
%
\institute{Amherst Center for Fundamental Interactions \\ Department of Physics, University of Massachusetts, Amherst\\ Amherst, MA 01003 USA}
\date{Received: date / Revised version: date}
%
\abstract{Both gravitons and pions are described by non-linear and non-renormalizable actions at low energies. These are most usefully treated by effective field theory, which is a full quantum field theoretic approach that relies only on the low energy degrees of freedom and their interactions. The gravitational case is particularly clean because of the masslessness of the graviton and the wide separation of scales. This essay provides an overview of this approach.
} 
\maketitle
%

\section{Introduction}\label{intro}

There is a story/parable attributed to the great particle theorist James Bjorken (aka BJ) that goes roughly as follows: In the old days theorists were generally humble. We would propose a new modification to present theory or a prediction based on our favorite model. Because we were aware that experimentalists were able to test this hypothesis very quickly, we did not attach ourselves to the predictions too highly - they might be immediately be shown to be wrong. If the idea was wrong we would drop it, if correct we would follow up. In the present days, there are many theoretical modifications to present theory that we all know will never be tested in our lifetimes. Yet the proponents are generally committed to working on these ideas for the rest of their careers. That requires a deep personal commitment. One comes to believe that one's own approach is the most promising, and that those of others are less valid. It develops a style like religion - one has to believe in what one is doing in order to continue spending your career on it. That is why the fights between the different approaches are so bitter - they are like religious wars.

If you think about it, this is a striking rebuke to much of what we do in philosophy of/and science. We often think ``what is the nature of reality?'' or ``what is the theory of everything?''. Yet BJ emphasizes that the question  is really ``what is the next step in our understanding of reality `now'?" The qualifier `now' means primarily the experimental frontier but also the frontier of our understanding of techniques - the frontiers of energy, emergence, understanding. We have gotten away from thinking of these frontiers as the limits of our knowledge and changed to the remarkable hubris that we can hope to solve everything without further experimental input.

But there is also a different take-away message to this story that I wish to emphasize here, and that is about the {\em tools} that we use in dealing with our theories. Back in the ``old days'' we had primarily two types of tools - models and full quantum field theories (designed to be valid at all scales). For some problems - such as nuclear physics or low-energy hadronic physics - neither of these is very good. The correct full quantum field theory - QCD - is not analytically tractable at low energies, and other possible renormalizeable field theories just do not match up with the complexity of the phenomena. Using phenomenological models was frustrating because they were uncontrolled and suffered from adjustable assumptions. So one of the reasons that we were humble in the old days is that we did not have the tools to trust our predictions\footnote{I am being somewhat unfair here. There was also a tradition of using dispersive techniques which in principle are fully rigorous. But in practice one had to make various approximations in applying dispersion relations to complex problems, and so also here the reliability was suspect. Dispersive techniques have been married to effective field theory successfully enhancing the power of both.}.

This changed when effective field theory came around. With effective field theory we could start making controlled and rigorous predictions. We could use only the degrees of freedom that we knew experimentally, and the symmetries that we had uncovered, and use these to make predictions which would hold independent of what the right QFT was at higher energies. Yes, there were many more unknown parameters, but that was OK because we had experiments to measure them. We theorists could be a little less humble and stand behind our predictions. Yet we could only this by embracing our ignorance of physics beyond the known frontiers.

This essay is a contribution to a volume on the relation of effective field theory (and the philosophy thereof) to the practice of nuclear physics. For the reasons mentioned above, the use of effective field theory was first adopted enthusiastically by the fields of low energy hadronic physics and nuclear physics. In some ways, the development of chiral perturbation theory is a good way to learn about the practical uses of effective field theory. I can always recommend my favorite text \cite{Donoghue:1992dd} as a good place to read about this. However, this essay is partially about another effective field theory, which I will argue is purer and illustrates the concepts even better - that of quantum general relativity. This illustrates how we can proceed with limited knowledge, but that the nature of reality could in principle change if were were able to push forward the experimental frontier.  I then address a limit of nuclear physics which shares many of the nice features of the gravitational case.

\section{General comments}\label{comments}

We appear to have a closed self-consistent layer of reality in the Standard Model plus General Relativity. If we are to be humble, we have to admit that we do not know what comes next. We do not even know what are the right degrees of freedom - the basic building blocks. At the present energies, everything looks like we are dealing with quantum fields. But there is a cautionary tale from condensed matter. In matter, the fundamental physics is described by the interactions between the atoms. This can be described by a potiential which is always close to a harmonic oscillator at small fluctuations. In one dimension,
 \begin{eqnarray}
S &=& \int d t L\left[y_{i}, \dot{y}_{i}\right]=\int d t \sum_{i}\left[\frac{1}{2} m \dot{y}_{i}^{2}-V\left(y_{i}-y_{i-1}\right)\right] \nonumber  \\
&\approx& \int d t \sum_{i}\left[\frac{1}{2} m \dot{y}_{i}^{2}-\frac{1}{2} k\left(y_{i}-y_{i-1}\right)^{2}\right]
\end{eqnarray}
If you quantize this and take the continuum limit, you end up with phonons. These are described by a quantum field with the action
\begin{equation}
S=\int d x d t \frac{1}{2}\left[\frac{1}{v^{2}}\left(\frac{\partial \phi}{\partial t}\right)^{2}-\left(\frac{\partial \phi}{\partial x}\right)^{2}\right]=\int d x d t \frac{1}{2} \partial_{\mu} \phi \partial^{\mu} \phi
\end{equation}
We see that phonons behave as massless fields and have an emergent Lorentz-like symmetry. Neither the fields nor the symmetry are properties of the underlying system. These latter facts are instructive for model building. We cannot count on the fact that our degrees of freedom, their interactions, nor even their symmetries hold at all scales. In one of my favorite speculations about the structure of the ultimate theory - Holger Nielsen's Random Dynamics \cite{Nielsen:1983qj}- the idea that there are {\em all} possible types of fluctuations at the smallest scales. But the hypothesis is that the only type of fluctuations that can survive to long distances are those where the masslessness of the fluctuation are protected by a symmetry - gauge bosons and chiral fermions. These are the important features of our world. So even the symmetries of the Standard Model may be emergent.

One use of effective field theories is to test for unusual properties even before one can access the new degrees of freedom. In our phonon example, we would expect that the emergent Lorentz-like symmetry would be broken by non-harmonic forces. That is, if we expand
\begin{equation}
V([y_i]) =  \frac12 k \left(y_{i}-y_{i-1}\right)^{2}+\frac{1}{4} \lambda\left(y_{i}-y_{i-1}\right)^{4}+\ldots
\end{equation}
in the continuum limit the Lagrangian would obtain a term which breaks this symmetry
\begin{equation}
S=\int d x d t \frac{1}{2}\left[\frac{1}{v^{2}}\left(\frac{\partial \phi}{\partial t}\right)^{2}-\left(\frac{\partial \phi}{\partial x}\right)^{2}-\overline{\lambda}\left(\frac{\partial \phi}{\partial x}\right)^{4}\right] \ \ .
\end{equation}
Searching for such a new operator in the Lagrangian is one classic use of effective field theories.

However, effective field theory is not just the use of effective Lagrangians - it involves the full use of quantum field theory including loop diagrams. Indeed this is its most interesting use. You can even do this in cases where you do not know the full theory or do not have a renormalizeable description. It answers a fundamental apparent problem in quantum physics. In perturbation theory you are instructed to sum over all intermediate states
\begin{equation}
\sum_{I} \frac{<f|V| I><I|V| i>}{E_{i}-E_{I}}
\end{equation}
including those which have not been discovered yet, and including possible interactions at the highest energies. How can we do such a calculation if we have no experimental information about these interactions? The solution to this problem involves the uncertainty principle. Although we do not know the high energy degrees of freedom or their interactions, we do know that they do not propagate far at low energy, so that their effect is basically local. By populating the effective Lagrangian with all possible local terms, with coefficients that are a-priori unknown, we would encompass the residual high energy effects. And most of these effective Lagrangians are not very relevant. They are suppressed by many powers of the low energy scale divided by the high energy scale and have effects which are numerically smaller than whatever precision we are working towards.

But the quantum physics of the light degrees of physics is not local, and is cannot be modified by any unknown physics that happens at high energy\footnote{Of course, effective field theory can also be used as a simplification of a known theory. The same rules apply, but the coefficients in this case can be known from matching to the full theory.}. But the key feature is that the quantum effects from the low energy parts of loops are reliable predictions of the effective field theory.

The above is just a very brief statement about the nature of an effective field theory. Some people also put a premium on naturalness, the requirement that the size of the corrections from higher order Lagrangians match some expected size - for example the size arising from loop effects. However, this is not a requirement for the existence of the effective field theory. The EFT accepts whatever parameters experiment measures. Moreover, the real quantum predictions of the EFT are independent of the parameters. The range of utility of the theory may depend on the parameters, but the existence of the EFT is independent of them. I will illustrate this in the discussion of the gravitational effective field theory in the next section.

\section{Gravity}\label{gravity}

General Relativity is an ideal case for the use of effective field theory. The theory contains gravitational waves, and quantum mechanics tells us that these imply massless gravitons as degrees of freedom. It is not hard to quantize General Relativity - it was done correctly by Feynman \cite{Feynman:1963ax}  and DeWitt \cite{DeWitt:1967ub}. Propagators and gravitational vertices are well defined.

This is not the place to get too technical about the gravitational effective field theory, but in fact the basic features can be simply understood. The action which leads to the Einstein equations is notationally simple - it just involves the curvature scalar
\begin{equation}\label{EH}
S_{g r a v}=\int d^{4} x \sqrt{-g}\left[\frac{2}{\kappa^{2}} R\right]
\end{equation}
with the identification $\kappa^2 =32\pi G$ with $G$ being Newton's constant. However the keys to unpacking this simplified notation is to note that the field involved is the metric tensor $g_{\mu\nu}(x)$, and the curvatures are all second order in derivatives in the metric. Specifically,
\begin{eqnarray}
 R_{\mu \nu} &=\partial_{\nu} \Gamma_{\mu \lambda}^{\lambda}-\partial_{\lambda} \Gamma_{\mu \nu}^{\lambda}+\Gamma_{\mu \lambda}^{\sigma} \Gamma_{\nu \sigma}^{\lambda}-\Gamma_{\mu \nu}^{\sigma} \Gamma_{\lambda \sigma}^{\lambda} \\ \Gamma_{\alpha \beta}^{\lambda} &=\frac{g^{\lambda \sigma}}{2}\left(\partial_{\alpha} g_{\beta \sigma}+\partial_{\beta} g_{\alpha \sigma}-\partial_{\sigma} g_{\alpha \beta}\right)
\end{eqnarray}
plus $R=g^{\mu\nu}R_{\mu\nu}$.

All of the tensor indices in these equations are just distractions from the physics. The only key point for us is to follow the derivatives. When matrix elements are taken, a derivative becomes a factor of an energy. So this action is of order $E^2$ or $\partial^2$. For the kinetic energy terms, this is similar to the Lagrangians of other massless fields. In addition it implies that the interaction terms in the Lagrangian are of order $E^2$.

If you look past all the tensor indices, the structure of GR is quite simple. If we look at the gravitational field $h_{\mu\nu}$ close to flat spacetime, with $g_{\mu\nu} = \eta_{\mu\nu} + \kappa h_{\mu\nu}$, the Lagrangian would become (schematically)
\beq\label{powers}
{\cal L}= \frac12 \partial h \partial h + \kappa h \partial h \partial h + \kappa^2 h^2 \partial h \partial h +....
\eeq
where the ellipses refer to yet higher powers of the field $h$. However, here we also see the potentially scary part for a quantum theory. The interaction terms are of the form that is labeled ``non-renormalizeable''. This is a misnomer, and the theory can be renormalized order by order in perturbation theory. However, in the old days (pre-EFT) this fact caused a lot of hand-wringing about the supposed incompatibility of quantum theory and general relativity. The reader has certainly seen ominous quotes to this effect - they still are seen even in the post-EFT world. Even though I once shared this concern, in retrospect it is puzzling to think why it got phrased this way. Yes, everyone was aware that these theories had extra divergences and general relativity would need to be changed in some way if it was to become renormalizeable. But that is not the same thing as a fundamental incompatibility with quantum mechanics.

In fact, Feynman first solved the only technical problem in quantizing GR in 1963 \cite{Feynman:1963ax}. That problem is that the metric tensor has more components than does the physical graviton degrees of freedom (two). Ever after gauge fixing, these extra components give extra spurious contributions in loops. The solution was to add an ``extra dopey particle'' (Feynman's phrase) with a minus sign to cancel off the spurious effects. This was formalized by DeWitt \cite{DeWitt:1967ub} and these are the Feynman-DeWitt-Fadeev-Popov ghosts, and the procedure is familiar to anyone who has gone through the quantization of QCD \footnote{For pedagogic treatments of this I recommend my lecture notes written up in Ref. \cite{Donoghue:2017pgk} and those of Veltman \cite{Veltman:1975vx}}. I often joke that the quantization of GR can be given as a homework exercise in a QFT class that has just quantized QCD, although I have never had the courage to actually do that. But quantization itself is not the problem.

Renormalization is also not a problem, although here we start to move into EFT territory. Early works in gravity QFT tend to focus on the divergences. These are high energy effects, and so they are equivalent to local terms in the Lagrangian. There are techniques, such as the background field method, that shows that they respect the symmetries of GR. But they are not just a renormalization of the Einstein-Hilbert action, Eq. \ref{EH}. One loop divergences occur at order of the curvature-squared \cite{tHooft:1974toh}, two loop at order curvature-cubed \cite{Goroff:1985sz}, etc. This power counting follow straightforwardly from the two derivative nature of the interaction terms, or equivalently to the dimensional coupling constant $\kappa$. However in a generally covariant action, these terms are not forbidden. If you include them in the action, the coefficients of the various terms can be used to renormalize all the divergences. These divergences are probably spurious, because the ultimate high energy theory is probably different from low energy GR, but that does not matter. There are probably also finite effects from the ultimate high energy theory - these also do not matter. All we need to know is that there is some parameter in a local action which encompasses the effects from the unknown high energy theory and which in principle can be measured by experiment.

Predictability was thought to be a problem. With new parameters popping up at each loop order, how could you make any predictions? Here is where EFT really comes to the rescue in a more significant way \cite{Donoghue:1994dn}. It shifts the focus from the UV, where we do not have any clue as to the ultimate theory, to the infrared, where we know gravitons and their interactions. The key here is non-locality - low energy gravitons can propagate long distances, and this is distinct from any local term in the action. In momentum space this manifests itself as non-analytic terms, powers of  $\sqrt{q^2}$ and $\log q^2$. Local terms in the action are simple powers of derivatives, corresponding to analytic terms which cannot be confused with the non-analytic counterparts. The non-local/non-analytic terms can be reliable predictions of the low energy part of the theory because they cannot be modified by any change in the theory that is made at high energy.

We can see the EFT predictability in action by looking at the results of some of the calculations. For example, there is the quantum correction to the Newtonian potential, which with an appropriate definition is found \cite{BjerrumBohr:2002kt}\cite{Khriplovich:2002bt} to be the final term in the expression
\begin{equation}
V(r)=-\frac{G m_{1} m_{2}}{r}\left[1+3 \frac{G\left(m_{1}+m_{2}\right)}{r}+\frac{41}{10 \pi} \frac{G \hbar}{r^{2}}\right] \ \  .
\end{equation}
The middle term is a classical correction due to GR. The important point is that the quantum correction is finite and independent of any parameters. The local terms in the action would yield a $\delta^3(x)$ correction to the potential in this particular case. The quantum correction therefore cannot be modified by anything that is done at high energy - it is a low energy theorem of quantum gravity. Moreover it is universal  \cite{Bjerrum-Bohr:2013bxa} - independent of the nature of the scattering particles.

Another similar prediction
is the bending angle of light or other massless particles in the presence of a large mass, which is found to be \cite{Bjerrum-Bohr:2014zsa}\cite{Bjerrum-Bohr:2016hpa}\cite{Bai:2016ivl}\cite{Chi:2019owc}
\begin{equation}
\theta =\frac{4 G M}{b}+\frac{15}{4} \frac{G^{2} M^{2} \pi}{b^{2}}+\frac{8 \mathrm{bu}^{\eta}-47+64 \log \frac{2 r_{0}}{b}}{\pi} \frac{G^{2} \hbar M}{b^{3}}
\end{equation}
where $b$ is the impact parameter. Again, none of the parameters from the local Lagrangian enter the result. Here, $r_0$ is related to an infrared divergence due to IR radiation. The bending angle is not universal, as the coefficient  $ \mathrm{bu}^{\eta}$ takes on different values for a photon, a massless scalar and a graviton, with $\mathrm{bu}^{\eta} = (371/120,~ 113/120, ~-29/8)$ for $\eta = $ (scalar, photon, graviton) respectively.  But including this spin dependence, this is another low energy theorem of quantum gravity. One consequence of this is that in quantum gravity there is no longer the concept of a lightcone nor universal geodesics in the presence of matter. This is possible because of the non-locality of the quantum correction - the quantum physics samples more than just the region of the classical trajectory.

These particular examples were chosen because it was clear (to an effective field theorist) that they would be independent of the local parameters. Other calculations may not be as fortunate. But still, even in such cases there are relations between reactions which would be independent of the parameters. In practice, for gravity these quantum effects are far too small to be measured. Depending on what one is doing, they are of order $10^{-50}\to 10^{-100}$ compared to the leading term. But we can turn this into a positive. Perturbation theory works best when the corrections are small. Quantum gravity is not the worst quantum theory as we used to think, but the best perturbative theory ever seen.

These examples highlight the real contribution of the effective field theory. It is not the divergences, nor the parameters, but only the low energy propagation, which is the real content of the theory.

 This also tells us that naturalness is not a requirement of the EFT. In this case, naturalness would refer to the coefficients of the terms higher order in the curvature, for example $c_1R^2$. The parameter $c_1$ is dimensionless, and naturalness would have it of order unity. In this case the energy expansion would work up to close to the Planck scale, and the effective field theory would potentially be valid to this scale. But if something drastic happened at lower energies, such as new dimensions opening up just beyond the weak scale, the coefficient could be very much larger because the EFT would break down at this much lower scale. It does not matter - the EFT will take whatever Nature tells us. But still the low energy predictions stand, independent of whether the coefficients are natural or not.

Of course, the effective field theory does not solve all the problems of quantum gravity. The effective field theory falls apart at or below the Planck scale. We do not have any experimental clue about what comes next - not even if it is a quantum field theory. It is nevertheless a fascinating theoretical area because we only vaguely understand the possibilities for a UV completionn for gravity and it is fun to explore the possibilities. But the field is not likely to become an experimental field.

The effective field theory has taught us a very important lesson: It is not an incompatibility of quantum theory with general relativity that we are confronted with, but rather a need to uncover the UV completion for gravity. Given the experimental realities, the EFT may be all that we can hope for in our lifetimes.

Since I opened this essay with a story from Bjorken, let me also use a quote from him to summarize this section.

{\it I also question the assertion that we presently have no quantum field theory of
gravitation. It is true that there is no closed, internally consistent theory of quantum
gravity valid at all distance scales, But such theories are hard to come by, and in
any case, are not very relevant in practice. But as an open theory, quantum gravity
is arguably our best quantum field theory, not the worst. Feynman rules for interaction
of spin-two gravitons have been written down, and the tree-diagrams (no closed
loops) provide an accurate description of physical phenomena at all distance scales between cosmological scales, down to near the Planck scale of $10^{{\rm -}33}$ cm. The divergent
loop diagrams can be renormalized at the expense of an in-principle infinite number
of counterterms appended to the Einstein-Hilbert action. However their effects are
demonstrably small until one probes phenomena at the Planck scale of distances and
energies

One way of characterizing the success of a theory is in terms of bandwidth, defined
as the number of powers of ten over which the theory is credible to a majority
of theorists (not necessarily the same as the domain over which the theory has been
experimentally tested). From this viewpoint, quantum gravity, when treated|as described above|as an effective feld theory, has the largest bandwidth; it is credible
over 60 orders of magnitude, from the cosmological to the Planck scale of distances.} \cite{Bjorken:2000zz}
\begin{figure*}
\resizebox{0.95\textwidth}{!}{%
  \includegraphics{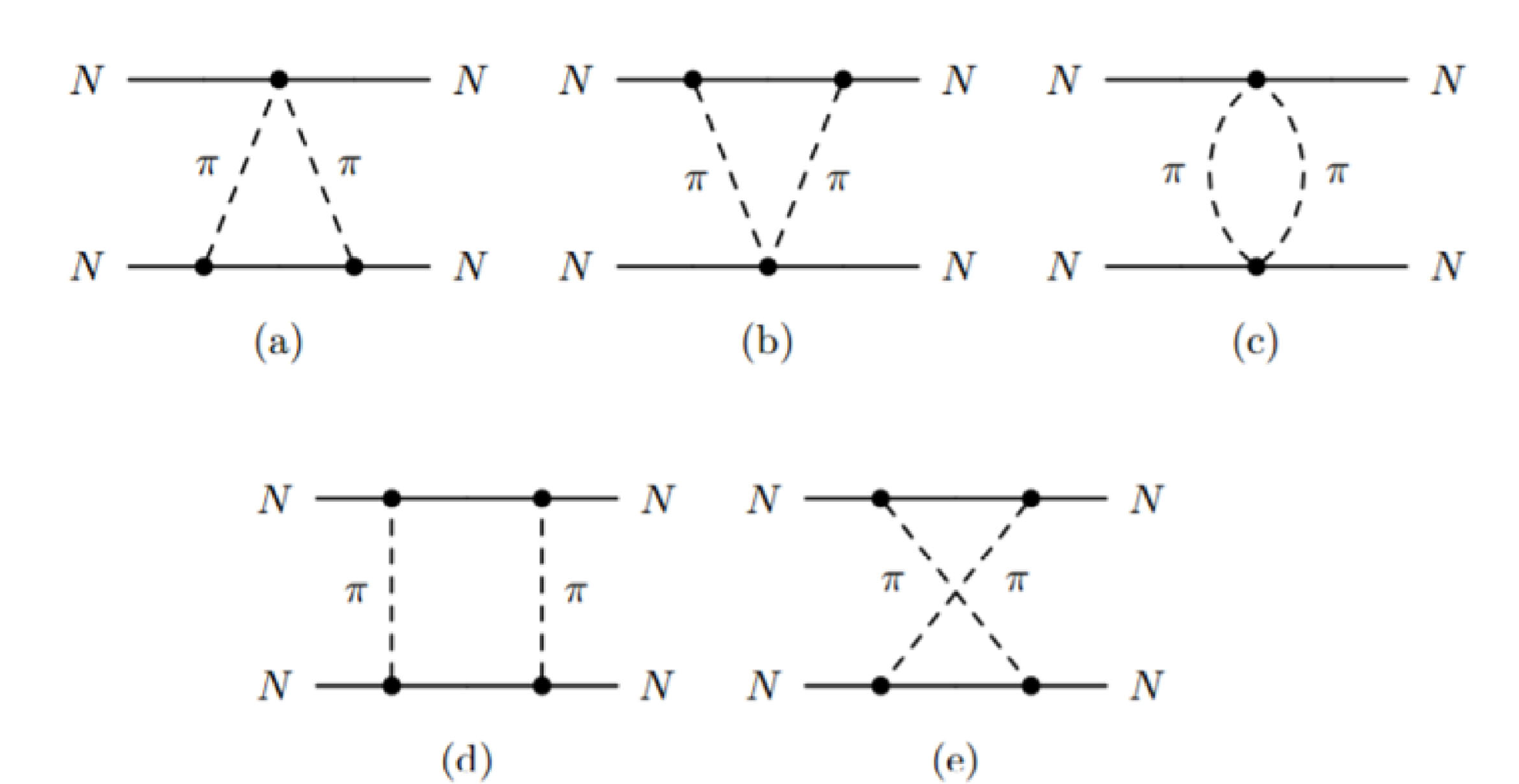}
}
\caption{The Feynman diagrams of the nucleon potential due to two pion exchange.}
\label{potentialfig}       
\end{figure*}

\section{Nuclear Physics}\label{nuclear}

The EFT of gravity is exceptionally ``clean'' because the graviton is massless. For nuclear physics, the pion mass introduces an additional scale, besides the QCD scale. In fact, the mass happens to have a value which, along with the values of the various coupling constants, introduces more scales - the binding energy per nucleon, the average nucleon momentum, the energy level spacings. This complicates the separation of scales in nuclei. That is why this conference/volume invokes a tower of effective field theories\footnote{As a sociological observation related to the BJ story at the start of this essay, I would also note that the differences between these variants of nuclear EFTs seemed to bring out the religious fervor of many participants.}. The interplay of these scales is very subtle and is responsible for the variety of nuclear structure. It may even have anthropic implications \cite{Damour:2007uv}\cite{Donoghue:2016tjk}.

It is interesting to consider an imaginary world where pions were strictly massless. In this world, there would be only one scale in nuclei and nuclear physics would relatively clean (up to the inclusion of QED and the electron mass). I wonder if this might be a useful starting point for a treatment of nuclear physics.


%
\label{potential}

If the pions were strictly massless they would play a role in everyday life. The Sun would emit them and they would be almost as ubiquitous as light. Condensed matter systems and atomic physics would also involve pions. Classical physics would have identified the currents involving pions, similar to the electromagentic current. We would have been searching for a reason for the masslessness of the pions and probably would have invoked a symmetry. It is even possible that someone clever (Albert Pionstein?) would propose a non-linear effective Lagrangian, which is the chiral Lagrangian,
\begin{equation}
\mathcal{L}=\frac{F^{2}}{4} {\rm Tr}\left(\partial_{\mu} U \partial^{\mu} U^{\dagger}\right)
\end{equation}
with
\begin{equation}
U=\exp \left[\frac{i \tau \cdot \pi}{F}\right]  \ \ .
\end{equation}
This would set off a long discussion of deep reasons why quantum mechanics and pion physics were incompatible. Finally EFT would come to the rescue and we could make clean predictions for the nuclear potential.

The calculation of the internucleon potential would be relatively straightforward in the EFT. Because the pion has an isospin quantum number there are several isospin components to the potential, i.e.
\begin{equation}
V(r)=V_{S}(r)+V_{T}(r) \tau_{1} \cdot \tau_{2}\left(\sigma_{1} \cdot \hat{r} \sigma_{1} \cdot \hat{r}-\frac{1}{3} \sigma_{1} \cdot \sigma_{2}\right)+\ldots . \ \ .
\end{equation}
The most important one for nuclear physics is the isoscalar and spin-scalar potential $V_S(r)$ which provides much of the binding force. A calculation, see Fig. \ref{potentialfig}, tells us that this has the form
\beq
V_{S}(r)=\frac{9}{16 \pi^{2} F_{\pi}^{2}}\left[\frac{g_{A}^{2} c_{3}}{r^{6}}-\frac{2 \overline{c}^{2}}{\pi r^{7}}\right]  \ \ .
\eeq
The parameters $F_\pi$ and $g_A$ are well known. There are also two pion couplings, $c_i$, which influence higher order corrections \footnote{The gravitational effective field theory has fewer parameters than the pionic one, because gravitons couple to the energy momentum tensor, which is independently known.}.  There are various conventions in the literature, and I am using those of Epelbaum et al. \cite{Epelbaum:2002gb}\cite{Epelbaum:2003gr}, see also \cite{Ordonez:1995rz}\cite{Beane:2002xf}. Here
\beq
\overline{c}^{2}=\left[c_{3}+\frac{c_{2}}{6}\right]^{2}+\frac{c_{2}^{2}}{45}
\eeq
Likewise the isospin tensor contribution is
\beq
V_{T}(r)=\frac{1}{16 \pi^{4}} \frac{g_{A}^{2}}{F_{\pi}^{2}} \frac{1}{r^{3}}\left[1-\frac{c_{4}}{\pi F_{\pi}^{2}} \frac{1}{r^{3}}\right] \ \ .
\eeq

The numbers for the various parameters in these potentials can be extracted from experiment. Those which we obtain in the real world have some contamination from the pion mass. But for our purposes, we can use these numbers. The scientists in this imaginary world will have discovered a new scale - the QCD scale. Numerically these potentials become
\begin{equation}
V_{S}(r)=-300~ \mathrm{MeV}~\left[\left(\frac{r_{0}}{r}\right)^{6}+0.24\left(\frac{r_{0}}{r}\right)^{7}\right]
\end{equation}
with
\begin{equation}
r_{0}=1~ \mathrm{fm}=(200 ~\mathrm{MeV})^{-1}   \ \ .
\end{equation}
Similarly
\begin{equation}
V_{T}(r)=-29 ~\mathrm{MeV}~\left[\left(\frac{r_{0}}{r}\right)^{3}-0.66\left(\frac{r_{0}}{r}\right)^{6}\right] \ \ .
\end{equation}
The QCD scale plays a role for these potentials that the Planck scale played for gravity.

In practice these potentials need to be supplemented by contact interactions, such as $C_S \bar{\psi}\psi\bar{\psi}\psi$, which carry coefficients (here $C_S$) which encode the residual information from the full theory. When dealing with potentials which are singular at $r\to 0$, there would need to be a regularization scheme to deal with the short distance divergences, and these regularization effects would appear in the renormalized values of the coefficients \cite{Long:2007vp}. But the basic point is that the structure of the long distance internucleon potential can be well understood in this limit.

\begin{figure}
\resizebox{0.5\textwidth}{!}{%
  \includegraphics{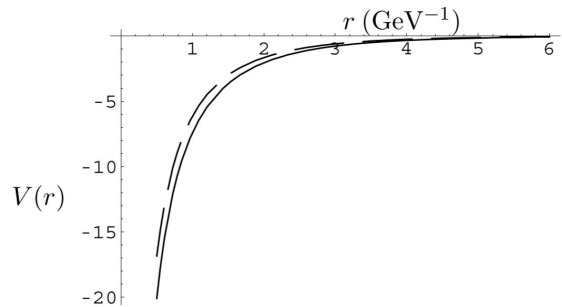}
}
\caption{The comparison between the  scalar-isoscalar potential in the chiral limit and a potential describing sigma exchange. }
\label{scalar}
\end{figure}

Despite the fact that this hypothetical world looks somewhat distant from the real world, this might be an interesting starting point for nuclear calculations. I have spent some time exploring the nuclear force in the limit of massless pions \cite{Donoghue:2006du}\cite{Donoghue:2006rg} and was struck by how much the central potential in the chiral limit resembles that of sigma exchange, which is often used to model this component in nuclear potentials. The comparison is shown in Fig. \ref{scalar}. This is important because sigma exchange is the least reliable aspect of the theory of nuclear potentials. The sigma appears not to be a fundamental state in QCD, but is generated as a strong-coupling effect far from the real axis in pion-pion scattering \cite{Leutwyler:2008xd}. Having a rigorous way to treat the two-pion channel near $m_\pi=0$, and then extrapolating towards the physical value, would provide insight

\begin{figure}
\resizebox{0.5\textwidth}{!}{%
  \includegraphics{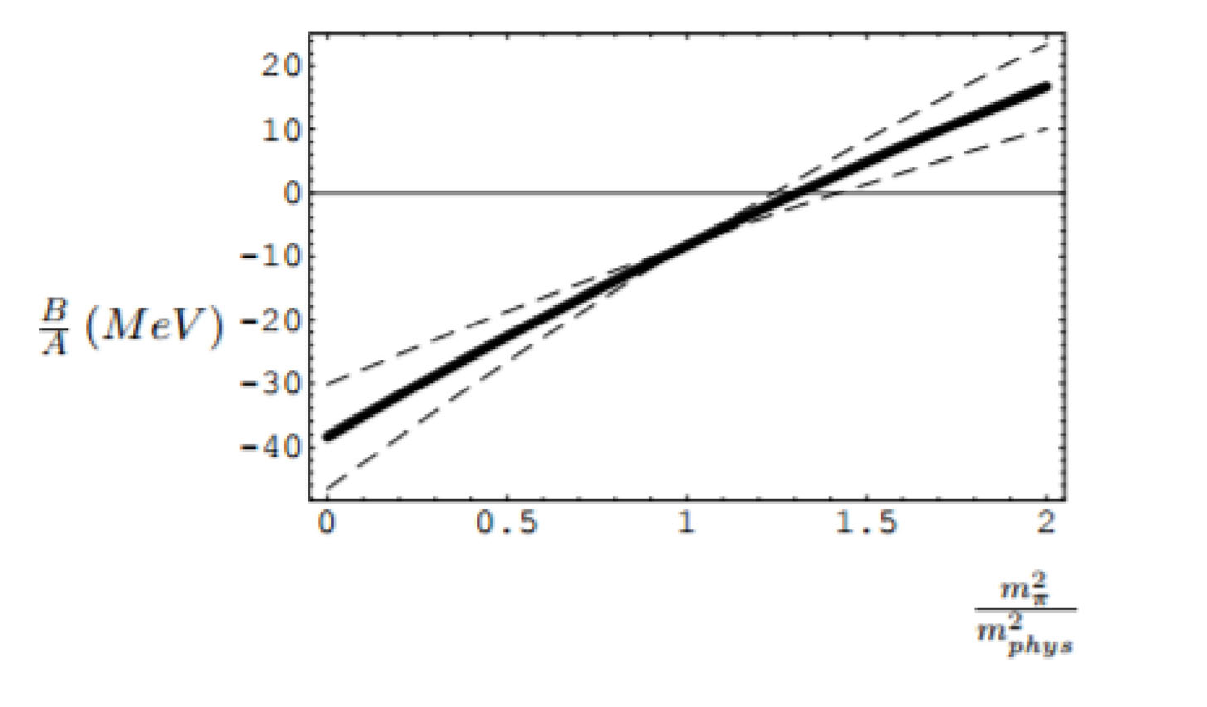}
}
\caption{The binding energy as a function of the pion mass. }
\label{binding}
\end{figure}

In these studies, I did not have the tools to provide a fully many-body treatment of nuclei, and instead used a different variant of effective field theory to convert the nuclear potential to binding energies \cite{Furnstahl:1999rm}. While in retrospect, I might have done the matching between the two effective field theories slightly differently, this also illustrates the power of effective field theory. The binding energy can be described by pion physics and a few extra low energy parameters. Calculating the mass dependence of the parameters allows on to bypass the complicated many-body calculations. In this approach, the binding energy per nucleon does have significant change as one heads to the chiral limit. This is shown in Fig. \ref{binding}.

The utility of this approach would depend on the ability to interpolate the results between the chiral limit and the physical pion mass. A lot of this dependence is kinematic - calculable dependence on the pion mass in propagators. The hidden dependence in the mass dependence of low energy parameters such as $g_A$ is more difficult, but lattice techniques are getting much better at calculating the mass dependence in such quantities.

\section{Final comments}\label{final}

The techniques of theoretical physics are also able to advance. One very useful development has been that of effective field theory. This allows us to admit that we are ignorant of the physics which occurs at higher energy beyond our experimental frontier, yet still make real predictions using what we do know at low energy. Philosophically this is very important. We do not have to resolve deep questions such as ``what is the ultimate nature of reality''. We can just humbly do our job with the understanding that experiment has already taught us. This is useful in many areas, but in gravitational physics it is especially so. The experimental resolution of the ultimate nature of quantum gravity is likely beyond reach. But we do have a quantum theory of general relativity, valid at ordinary energies. This allows General Relativity to be included in our present Core Theory along with the Standard Model.

\section*{Acknowledgements} I would like to thank the organizers and participants at the workshop ``The tower of the effective field theories and the emergence of the nuclear phenomena'' (January 2017, CEA Saclay) for lively discussions, most particularly U. Meissner, U. van Kolck and P. Vanhove. I also thank J. Bjorken for confirming the story used in the introduction. This work has been supported in part by the National Science Foundation under grant NSF PHY15-20292 and PHY18-20675

\end{document}